\documentclass[prb,preprint,groupedaddress,english]{revtex4-1}

\usepackage{amsmath}
\usepackage{bm}
\usepackage{graphicx}
\usepackage{nicefrac}

\newcommand{\matrixel}[3]{\left< #1 \vphantom{#2#3} \right| #2 \left| #3 \vphantom{#1#2} \right>}

\begin{document}
\title{Uniform Electron Gas under An External Bias: The Generalized Thomas-Fermi-Dirac Model and the Dual-Mean-Field Theory}


\author{Chun Zhang$^{1,2}$}
\email{phyzc@nus.edu.sg}
\affiliation{
	$^1$Department of Physics, National University of Singapore,
		2 Science Drive 3, Singapore 117542\\
	$^2$Department of Chemistry, National University of Singapore,
	  3 Science Drive 3, Singapore 117543}

\begin{abstract}
The uniform electron gas placed between two reservoirs is used as a model system for molecular junctions under an external bias. The energetics of the electron gas are calculated by generalizing the Thomas-Fermi-Dirac (TFD) model to nonequilibrium cases. We show that when the bias voltage is not zero, the first Hohenberg-Kohn (HK) theorem breaks down, and energies of the electron gas can be determined by the total electron density together with the density of nonequilibrium electrons, supporting the dual-mean-field (DMF) theory recently proposed by us [J. Chem. Phys. 139, (2013) 191103]. The generalization of TFD functionals to DMF ones is also discussed.  
\end{abstract}


\maketitle


Building electronic devices on the basis of single molecules has drawn dramatic attention in past two decades.~\cite{AR_diode} On the theoretical side, the \textit{ab initio} method that combines the density functional theory (DFT) and nonequilibrium Green's functions' (NEGF) techniques has proven to be powerful in describing electron quantum transport at molecular scale,~\cite{GHPRB} and has achieved great success in understanding and designing molecular-scale electronic devices.~\cite{Ratnerreview, polymerjcp, zhangcswitch, CZ, ZX, ZM, ZAH, CYQ} In this method, the mean-field potential obtained from DFT is used to calculate electronic structures of molecular junctions with or without an external bias. In our recent paper~\cite{DMF}, we have shown that when a finite external bias is present, the properties of a molecular junction can not be determined by the total electron density alone, and a dual-mean-field (DMF) theory is proposed to incorporate the bias-induced nonequilibrium effects. One of the key results of the DMF theory is that the current-carrying electrons experience a different effective mean-field potential from the equilibrium electrons do. In this paper, we present detailed analysis of energetics of the uniform electron gas (UEG) placed between two reservoirs. Our calculations clearly show that the first Hohenberg-Kohn (HK) theorem that is the basis of DFT breaks down for the system under study when the external bias is not zero, providing a strong support for the DMF theory~\cite{DMF}.


\begin{figure}
\includegraphics[]{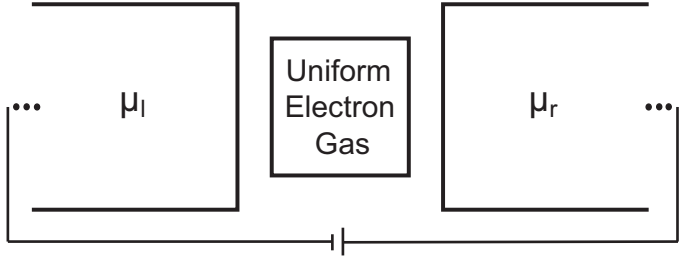}
\caption{ The uniform electron gas is sandwiched by two reservoirs. A battery maintains the chemical potentials, $\mu_l$ and $\mu_r$, in the left and right reservoirs, respectively. The bias voltage acrossing the uniform electron gas is defined by $(\mu_l-\mu_r)/e$. We assume $\mu_l > \mu_r$.}
\end{figure}

The model is shown in Fig. 1, where an uniform electron gas is placed between two reservoirs. A battery is connected to the system to maintain the chemical potentials $\mu_l$ and $\mu_r$ for the left and right reservoirs respectively. The bias voltage accrossing the system can then be defined as $V_b =(\mu_l-\mu_r)/e$. When $V_b$ is not zero, the uniform electron gas is in a nonequilibrium state. The nonequilibrium distribution is as follows: Electrons coming from the left reservoir obeys the distribution $f_{FD}(\mu_l)$, and electrons coming from the right reservoir obeys the distribution $f_{FD}(\mu_r)$, where $f_{FD}$ is the equilibrium Fermi-Dirac distribution function. In Fig. 2, we schematically show the distribution in momentum space. 
By defining two Fermi vectors, $\frac{\hbar^{2}k_{fr}^{2}}{2m_{e}}=\mu_{r}$ and $\frac{\hbar^{2}k_{fl}^{2}}{2m_{e}}=\mu_{l}$, the kinetic and exchange energies of the uniform electron gas can then be analytically calculated in terms of two Fermi vectors by generalizing the Thomas-Fermi-Dirac (TFD) model to nonequilibrium cases~\cite{TFD}.

The kinetic energy density can be worked out as follows.
\begin{eqnarray}
k(k_{fr}, k_{fl})	&=& \frac{1}{V} \sum_{\boldsymbol{k},s} \matrixel{\boldsymbol{k},s}{\hat{t}}{\boldsymbol{k},s} \nonumber \\
	&=& \frac{1}{V} \sum_{\boldsymbol{k},s} \frac{|\boldsymbol{k}|^{2}}{2} \nonumber \\
	&=& \frac{1}{V} \sum_{\boldsymbol{k}} |\boldsymbol{k}|^{2} \nonumber \\
	&\stackrel{L \to \infty}{=}& \frac{1}{V} \int_{\Omega_{1} + \Omega_{2}} \frac{\mathrm{d}\boldsymbol{k}}{\left(\nicefrac{2\pi}{L}\right)^{3}} |\boldsymbol{k}|^{2} \nonumber \\
	&=& \frac{1}{10\pi^{2}} \cdot \frac{1}{2}(k_{fr}^{5} + k_{fl}^{5}).
\end{eqnarray}

\begin{figure}
\includegraphics[width=8.2cm]{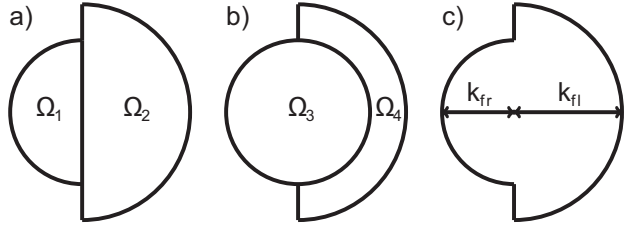}
\caption{ Non-equilibrium distribution in momentum space for the generalized Thomas-Fermi-Dirac model. a) The division of the momentum space into $\Omega_{1}$ and $\Omega_{2}$ for calculating the kinetic energy density. b) The division of the momentum space into $\Omega_{3}$ and $\Omega_{4}$ for calculating the exchange energy density. c) The momentum space is defined in terms of of two Fermi wavevectors $k_{fl}$ and $k_{fr}$.}
\end{figure}

The exchange energy density can also be analytically calculated as demonstrated in our previous paper~\cite{}. Here we show the final result.
\begin{eqnarray}
e_x(k_{fr}, k_{fl}) &=& -\frac{1}{16\pi^{3}} \bigg[ -\left(k_{fl}^{2} - k_{fr}^{2}\right)^2 ln\left(k_{fr}+k_{fl}\right) + k_{fr}^4  ln\left(k_{fr}\right) + \nonumber \\
	& & \left(k_{fl}^4 - 2k_{fr}^2 k_{fl}^{2}\right)ln\left(k_{fl}\right) + k_{fr}^4 + k_{fr}^3 k_{fl} \nonumber \\
	& & - \frac{1}{2}k_{fr}^2 k_{fl}^2 + k_{fr} k_{fl}^3 + \frac{3}{2} k_{fl}^4 \bigg].
\end{eqnarray}

Using the relationship  $\rho_{t} = \frac{1}{6\pi^{2}}\left(k_{fr}^{3}+k_{fl}^{3}\right)$ where $\rho_t$ is the total electron density, and also the definitions of two Fermi vectors, $\rho_t$ and the bias voltage $V_b$ can be calculated in terms of $k_{fr}$ and $k_{fl}$. Then the kinetic energy density (Eq. 1) and exchange energy density (Eq. 2) can be expressed as functions of $\rho_t$ and $V_b$. We plot the ratio of energy densities between the generalized TFD model and the equilibrium TFD model in Fig. 3a and 3b. Note that in equilibrium TFD model, $k^{TFD}(\rho_{t})=\frac{1}{10\pi^{2}}(3\pi^{2}\rho_{t})^{\nicefrac{5}{3}}$ and $e_{x}^{TFD}(\rho_{t})=-\frac{1}{4\pi^{3}} \left(3\pi^{2}\rho_{t}\right)^{\nicefrac{4}{3}}$. These figures (Fig. 3a, 3b) clearly suggest that the kinetic and exchange energies of the UEG under a finite bias cannot be uniquely determined by the total electron density alone, supporting the DMF theory~\cite{} we proposed. Note that at low biases around 1 V, the error of equilibrium TFD model can be very significant.   

\begin{figure}
\includegraphics[width=0.8\textwidth]{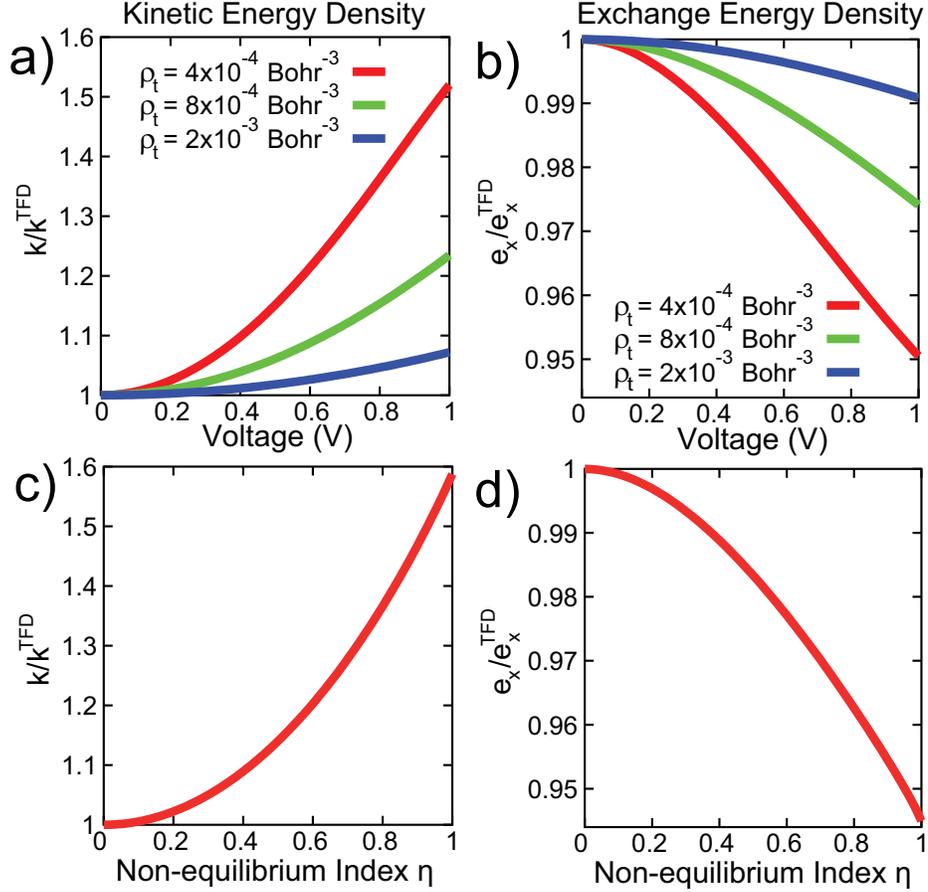}
\caption{ (color online) The kinetic and exchange energy densities of a uniform electron gas connected with two reservoirs. a) and b) show the ratio between non-equilibrium energy densities calculated from the generalized TFD model and the corresponding equilibrium ones. Clearly, the ratio depends on both electron density and the bias voltage. c) and d) show the correction factor as a function of non-equilibrium index $\eta$ for kinetic and exchange energy densities.}
\end{figure} 

Defining the nonequilirium electron density $\rho_{n} = \frac{1}{6\pi^{2}}\left(k_{fl}^{3}-k_{fr}^{3}\right)$ and the nonequilibrium index $\eta$ as the ratio between $\rho_{n}/\rho_{t}$, the kinetic and exchange energy densities can be rewritten in DMF format as Eq. 3 and 4. The DMF exchange energy density shown here (Eq. 4) is the same as the one presented in our previous paper~\cite{DMF}.
\begin{equation}
k(\rho_{t}, \eta) = k^{TFD}(\rho_{t}) \cdot \frac{1}{2} \left[ \left(1-\eta\right)^{\nicefrac{5}{3}} + \left(1+\eta\right)^{\nicefrac{5}{3}} \right],
\end{equation}
\begin{eqnarray}
e_x(\rho_{t}, \eta)	&=& e_{x}^{TFD}(\rho_{t}) \cdot \frac{1}{4}\left(1+\eta\right)^{\nicefrac{4}{3}} \bigg[ -\left(1-\tilde{\eta}^{2}\right)^{2}ln\left(1+\tilde{\eta}\right) \nonumber \\
	& & + \tilde{\eta}^{4}ln\left(\tilde{\eta}\right) + \tilde{\eta}^{4} + \tilde{\eta}^{3} - \frac{1}{2}\tilde{\eta}^{2} + \tilde{\eta} + \frac{3}{2} \bigg],
\end{eqnarray}

In above equations (Eq. 3 and 4),  $\tilde{\eta} \equiv \left(\frac{1-\eta}{1+\eta}\right)^{\nicefrac{1}{3}}$. Both DMF kinetic and exchange energy densities are in the format of the equilibrium TFD energy multiplied by a correction factor which is a function of $\eta$. When $\eta = 0$, DMF energy densities reduce to equilibrium TFD ones. When $\eta = 1$, the correction factor for the kinetic energy is around 1.6, and the correction factor for the exchange energy is around 0.95. The correction factor (the ratio between DMF energy and equilibrium TFD one) for the kinetic (exchange) energy density as a function of $\eta$ is shown in Fig. 3c (3d). 


In conclusion, in this paper, we present a detailed analysis for the energetics of the uniform electron gas under a finite bias. We show that when the bias is not zero, the properties of the electron gas cannot be uniquely determined by the total electron alone. The kinetic and exchange energy densities of the system are functions of total electron density and the bias voltage. We then define a so-called nonequilibrium electron density, and show that the energetics of the nonequilibrium uniform electron gas are functionals of the total electron density and the nonequilibrium electron density, supporting the DMF theory proposed in an earlier paper.

\vspace{5mm}
\noindent{\bf Acknowledgments.} We acknowledge the support from Ministry of Education (Singapore) and NUS academic research grants (R-144-000-325-112 and R144-000-298-112). Computations were performed
at the graphene research centre and centre for Computational Science and Engineering at NUS.



\begin{thebibliography}{99}

\bibitem{AR_diode}
A. Aviram, and M. A. Ratner, Chem. Phys. Lett. 29, (1974) 277.

\bibitem{GHPRB}
 J. Taylor, H. Guo, and J. Wang, Phys. Rev. B 63, (2001) R121104.
 
\bibitem{Ratnerreview}
S. M. Lindsay, and M. A. Ratner, Adv. Mater. 19 (2007) 23.

\bibitem{polymerjcp}
J. Tomfohr, and O. F. Sankey, J. Chem. Phys. 120 (2004) 1542. 

\bibitem{zhangcswitch}
C. Zhang, M. H. Du, H. Cheng, X. Zhang, A. Roitberg, and J. L. Krause, Phys. Rev. Lett. 92 (2004) 158301.

\bibitem{CZ}
C. Zhang, L. L. Wang, H. Cheng, X. Zhang, and Y. Xue, J. Chem. Phys. 124 (2006) 201107.

\bibitem{ZX}
Z. Dai, A. Nurbawono, A. Zhang, M. Zhou, Y. Feng, and C. Zhang, J. Chem. Phys. 134 (2011) 104706.

\bibitem{ZM}
M. Zhou, Y. Cai, M. Zeng, C. Zhang, and Y. Feng, Appl. Phys. Lett. 98 (2011) 143103.

\bibitem{ZAH}
A. Zhang, Y. Wu, S. Ke, Y. Feng, and C. Zhang, Appl. Phys. Lett. 22 (2011) 435702.

\bibitem{CYQ}
Y. Cai, A. Zhang, Y. Feng, and C. Zhang, J. Chem. Phys. 135 (2011) 184703.

\bibitem{DMF}
S. Liu, Y. Feng, and C. Zhang, J. Chem. Phys. 139 (2013) 191103.

\bibitem{TFD}
N. H. March, Adv. in Phys. 6 (1957) 1.


\end{thebibliography}
\end{document}